\documentclass[submission,copyright,creativecommons]{eptcs}
\providecommand{\event}{MeTRiD 2018} 
\usepackage{breakurl}             
\usepackage{underscore}           
\usepackage{color}
\usepackage{amsmath}
\usepackage{amssymb}
\usepackage{graphicx}
\usepackage{listings}
\usepackage{makecell}

\newcommand{\domanda}[1]{\paragraph{#1}\mbox{}\par\nobreak\noindent\ignorespaces}

\title{Model-Based Design of Energy-Efficient Applications for IoT Systems}
\author{Alexios Lekidis
\institute{Department of Informatics, \\ Aristotle University of Thessaloniki \\ 54124 Thessaloniki, Greece}
\email{alekidis@auth.gr}
\and
Panagiotis Katsaros
\institute{Department of Informatics, \\ Aristotle University of Thessaloniki \\ 54124 Thessaloniki, Greece}
\email{katsaros@csd.auth.gr}
}

\def\titlerunning{Model-Based Design of Energy-Efficient Applications for IoT Systems}
\def\authorrunning{A. Lekidis \& P. Katsaros}
\begin{document}
\maketitle

\begin{abstract}
A major challenge that is currently faced in the design of applications for the Internet of Things (IoT) concerns with the optimal use of available energy resources given the battery lifetime of the IoT devices. The challenge is derived from the heterogeneity of the devices, in terms of their hardware and the provided functionalities (e.g data processing/communication). In this paper, we propose a novel method for (i) characterizing the parameters that influence energy consumption and (ii) validating the energy consumption of IoT devices against the system's energy-efficiency requirements (e.g. lifetime). Our approach is based on energy-aware models of the IoT application's design in the BIP (Behavior, Interaction, Priority) component framework. This allows for a detailed formal representation of the system's behavior and its subsequent validation, thus providing feedback for enhancements in the pre-deployment or pre-production stages. We illustrate our approach through a Building Management System, using well-known IoT devices running the Contiki OS that communicate by diverse IoT protocols (e.g. CoAP, MQTT). The results allow to derive tight bounds for the energy consumption in various device functionalities, as well as to validate lifetime requirements through Statistical Model Checking.
\end{abstract}

\section{Introduction} \label{sec:intro}
The evolution of the IP-based internet towards the Internet of Things (IoT) has introduced innovations in applications from various domains, such as the smart grid, the building and home automation, the health monitoring and has even boosted a new industrial area, the so-called Industry 4.0 or Industrial IoT (IIoT) \footnote{\url{https://industrial-iot.com/2017/09/iiot-smart-manufacturing/}}. IoT leverages miniature devices that can exchange data autonomously through wireless communication. These devices, are usually of small size and low cost, and are also supplied with battery power, in order to widen the applicable deployment possibilities. An important challenge related to the use of battery power is that the device lifetime depends solely on the resource demands imposed by the IoT application.

The main difficulty in addressing this challenge is the underlying nature of IoT applications, namely that they are based on web services designed for continuous and long-lived service delivery through IoT devices with limited lifetime~\cite{colitti2011rest}. Thus, researchers have focused on mechanisms and protocols for low-power wireless communication, as well as on energy-efficient device hardware design \cite{benini1998system}. However, to the best of our knowledge, there is only limited work towards methods and techniques for real-time monitoring and characterization of the energy consumption in IoT devices~\cite{georgiou2017iot}. The main reason behind this is that such methods usually require direct interaction with the device hardware, which in most cases is not supported by the devices \cite{dunkels2007software}. Moreover, the existing analytical methods to estimate energy consumption in resource-constrained devices \cite{zhou2011modeling} use the device manufacturer characteristics, which may not be always accurate when compared with measurements taken in the system's operation environment \cite{vilajosana2014realistic}. 


To address these limitations of existing methods and techniques for monitoring the energy consumption of IoT devices, Dunkels et al. \cite{dunkels2007software} introduced a software-based solution, which is available for IoT applications in the Contiki OS through the powertrace module. Compared to a hardware-oriented approach, powertrace allows deriving a generic and hardware-agnostic model that can be applied to various device types, and enables a fine-grained analysis of the energy consumption at the network-level. However, powertrace can only support energy monitoring for the individual IoT devices, as well as the entire system. A resulting limitation is that powertrace cannot measure nor estimate the energy consumption for the device communication with its connected peripherals (e.g. sensors/actuators). A possible solution to this end should aim towards a proper characterization and assessment of all the parameters and scenarios that are impacting the energy consumption on a system-level, as described in \cite{martinez2015power}. Specifically, that work focuses on characterizing and estimating energy evolution over time in Wireless Sensor Networks (WSN) through distribution fitting techniques. Distribution fitting is a result of energy profiling in different WSN scenarios; the authors also stress that the analysis and correlation of parameters influencing energy consumption is a fundamental step towards a system design flow. 


In this paper, we propose a model-based characterization of energy consumption in IoT systems and in their constituent devices. This is achieved by the progressive development of faithful models at the system-level, which incorporate valid energy profiles for the various system operation scenarios. The models are implemented in the Behavior-Interaction-Priority (BIP) component framework \cite{basu2011rigorous} based on specifications for the functional behavior and energy constraints for the system under design. BIP enables effective semantics-preserving transformations for the system-level models, as well as the simulation and validation of IoT systems in every development stage. The approach supports the system's validation through Statistical Model Checking (SMC) against the related requirements provided as user input. This results in valuable feedback to the system designer for enhancements concerning: (i) the IoT device lifetime and (ii) the communication/computation energy cost in each IoT device, its sensors/actuators as well as in the overall system. 

The approach is illustrated through a Building Management System (BMS) which features several well-known IoT devices, such as the Zolertia Z1 \footnote{\url{http://zolertia.sourceforge.net/wiki/images/e/ec/Z1SP_Datasheet_v1.3.pdf}}, the Sky mote \footnote{\url{http://www.eecs.harvard.edu/~konrad/projects/shimmer/references/tmote-sky-datasheet.pdf}}, the OpenMote \footnote{\url{http://openmote.com/}} and SimpleLink Sensortag \footnote{\url{http://www.ti.com/ww/en/wireless_connectivity/sensortag/}}. To this respect, we provide detailed profiling and characterization of the energy consumption, as well as methods to build efficient IoT applications in BMS systems. Concretely, this paper has the following contributions:
\begin{itemize}
\item an energy-aware model that allows providing valid bounds for the energy consumption in different device functionalities (e.g. data processing/communication) 
\item a detailed analysis of the parameters that influence energy consumption in IoT systems, as well as the results from their integration into the energy-aware model
\item a validation technique for energy-consumption and battery lifetime requirements of an application design through SMC
\end{itemize}

The analysis of parameters that affect energy consumption through our approach allows for obtaining trustworthy design decisions concerning the scenarios that have a predominant effect in the energy efficiency of the IoT devices and the overall system.   



The rest of the paper is organized as follows. Section \ref{sec:context} provides a 
brief introduction in the Contiki IoT ecosystem and the techniques for software-based energy management through the powertrace, as well as in the 
BIP framework, which is used for rigorous system design. Section \ref{sec:flow} illustrates the proposed method for energy-consumption management 
in IoT systems, which is later used in Section \ref{sec:caseStudy} to validate IoT system requirements and provide valid bounds for the lifetime of IoT devices. Finally, Section \ref{sec:conc} provides conclusions and perspectives for future work.

\section{Background} \label{sec:context}
\subsection{Contiki powertrace} \label{sec:powertrace}

Powertrace \cite{dunkels2007software} is a Contiki library that allows the annotation of Contiki programs with primitives, for monitoring the energy flow in IoT devices. It identifies four individual operating modes that contribute to a device's energy consumption: 

\begin{itemize}
\item{Low Power (LPM)}: the device is idle waiting for an event
\item{CPU:} the device microcontroller is used for calculations/data processing
\item{Radio transmission (Tx)}: indicating data transmission
\item{Radio reception (Rx)}: indicating data reception 
\end{itemize}

The energy consumption varies according to the time that a device remains
in each of the above modes. In order to measure this time, the library provides code primitives that can be used for every IoT device type. A data logger is used to store the data, which supports energy analytics. Characteristic examples of such analytics are the duty cycle or the device lifetime. The former refers to the percentage of time that a device remains in one operating mode, whereas the latter refers to the total time duration that a device operates autonomously. The period that powertrace uses to measure and log the data can be configured by the user and has an impact on the performance and accuracy of the mechanism. Finally, the energy calculation in powertrace also supports hardware-specific parameters, such as the real-time timer (RTIMER \footnote{\url{http://anrg.usc.edu/contiki/index.php/Timers\#Step_5_-_Introduction_to_rtimer}}) that is used to measure the hardware clock cycles of the device per second. 




\subsection{The BIP component framework} \label{sec:smc-bip}

BIP  (Behavior-Interaction-Priority)  \cite{basu2011rigorous}  is  a  highly  expressive, component-based framework with rigorous semantic basis. It allows the construction of complex, hierarchically structured models from atomic components, which are characterized
by their behavior and interfaces. Such components are transition systems enriched with data. Transitions are used to move
from a source to a destination location. Each time a transition
is  taken,  component  data  (variables)  may  be  assigned  with
new  values,  which  are  computed  by  user-defined  functions
(in  C/C++).  Atomic  components  are  composed  by  layered
application  of  interactions  and  priorities.  Interactions  express
synchronization  constraints  and  define  the  transfer  of  data
between the interacting components. Priorities are used to filter
amongst possible interactions and to steer system evolution so
as to meet performance requirements, e.g. to express scheduling policies. A set of atomic components can be composed into
a generic compound component by the successive application
of connectors and priorities. 

BIP is supported by a rich toolset
including tools that are used to check stochastic systems, through the
\textit{Statistical Model Checking} (SMC) technique. SMC was proposed as a means
to cope with the scalability issues in numerical methods for the analysis of stochastic systems. Consider a system model $M$
and a set of requirements, 
where each requirement can be formalized by a stochastic temporal property $\phi$ written in the Probabilistic Bounded Linear Temporal Logic
(PBLTL)~\cite{herault2004approximate}. SMC applies a series of simulation-based analyses to decide PBLTL properties of the
following two types:
\vspace{-.1cm} 
\begin{enumerate}
\item \textit{Is the probability $Pr_M(\phi)$ for $M$ to satisfy $\phi$ greater or equal to a threshold $\theta$?} Existing approaches to answer this question are based on hypothesis testing~\cite{legay2010statistical}. When $p = Pr_M(\phi)$, to decide 
if $p \geq \theta$, we can test \textit{H}: $p \geq \theta$ against 
\textit{K}: $p < \theta$. Such a solution does not guarantee a correct 
result but it allows to bound the error probability. The strength of a test is determined by the parameters $(\alpha, \beta)$, such that the probability of accepting \textit{K} (resp. \textit{H}) 
when \textit{H} (resp. \textit{K}) holds is less than or equal to $\alpha$ (resp. $\beta$). 
However, it is not possible for the two hypotheses to hold simultaneously 
and therefore the ideal performance of a test is not guaranteed. 
A solution to this problem is to relax the test by working with an 
indifference region $(p_1$, $ p_0)$ with $p_0 \geq p_1$ ($p_0 - p_1$ is the 
size of the region). In this context, we test the hypothesis 
$H_0 : p \geq p_0$ against $H_1 : p \leq p_1$ instead of \textit{H} against 
\textit{K}. If the value of $p$ is between $p_1$ and $p_0$ (the indifference region), 
then we say that the probability is sufficiently close to $\theta$, so that we 
are indifferent with respect to which of the two hypotheses \textit{K} or \textit{H} is accepted.

\item \textit{What is the probability for $M$ to satisfy $\phi$?} This analysis computes the value of $Pr_M(\phi)$ that depends on the existence of a counterexample to $\neg \phi$, for the threshold $\theta$.  This computation is of polynomial complexity and, depending on the model $M$ and the property $\phi$, it may or may not terminate within a finite number of steps.  In~\cite{herault2004approximate}, a procedure based on the Chernoff-Hoeffding bound~\cite{hoeffding1963probability} was proposed, to compute a value for $p'$, such that $|p'-p|<\delta$ with confidence $1-\alpha$, where $\delta$ denotes the precision.
\end{enumerate}

The SMC of BIP models is automated by the SMC-BIP tool \cite{nouri2015statistical} that supports both types of PBLTL properties. The tool accepts as inputs the PBLTL property, a model in BIP and a couple of confidence parameters. The tool provides a verdict in the form of the probability for the property to hold true. Since the approach is designed for the validation of bounded LTL properties, it is guaranteed to terminate in finite time.

\section{Rigorous design of energy-efficient IoT systems} \label{sec:flow}
The overall flow of our method is presented in Figure \ref{fig:flow}. The process depends on an XML parameter configuration input file that extends the WPAN network configuration given in \cite{lekidis2015design} as well as on the application design expressed in a Domain Specific Language (DSL) \cite{lekidis2018model}. The third input is the requirement specification, which contains user requirements regarding energy constraints for the IoT application. The method proceeds throughout the steps described below:

\begin{figure}[htbp]
\begin{centering}
\includegraphics[width=0.504\textwidth]{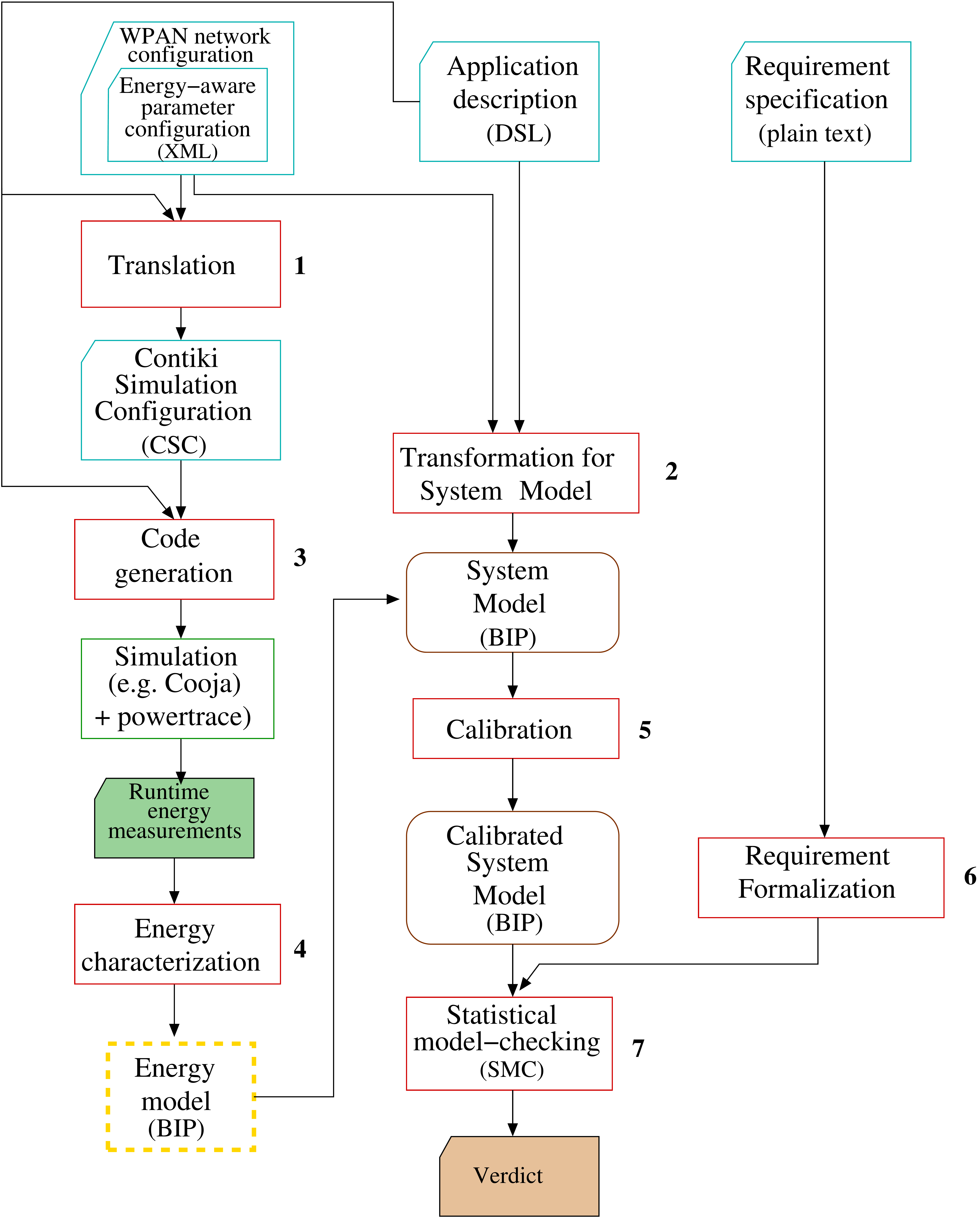}
\caption{The proposed design method}
\label{fig:flow}
\end{centering}
\end{figure}

\begin{enumerate}
\item{{\bf Translation for the generation of the Contiki Simulation Configuration}: This step leverages the XML-based configuration with parameters that affect the energy consumption in an IoT application, as they are presented in Section \ref{sec:energyParameters}. The configuration is systematically translated in order to produce all the necessary configuration files for the deployment of a Contiki IoT application as well as the Contiki Simulation Configuration (CSC) file, with the system architecture that is simulated within the native Contiki simulation environment (Cooja~\cite{eriksson2009cooja}). The DSL in \cite{lekidis2018model} is also used to provide all the necessary application-specific parameters during the translation stage.}
\item{{\bf Transformation for the System Model}}: The actions comprising this step are two-fold. First, the DSL application description of the previous step is used to form an Application Model, which is later enhanced with the OS/kernel model, that is formed from the BIP IoT component library \cite{lekidis2018model}. The combination of the two models is performed through the addition of the application mapping in the DSL~\cite{lekidis2018model}, that specifies the deployment of application modules onto the IoT system’s devices.
\item{{\bf Code generation}}: The CSC file from step 1 along with the DSL application description are used to generate deployable code. The code is annotated with energy characteristics to allow the calculation of the energy consumed in each operating mode of powertrace. The code is accordingly simulated in Cooja with the addition of the powertrace library. The simulation result is used for the energy characterization in the next step. 
\item{{\bf Energy characterization}}: The analysis of the energy-oriented behavior and characteristics leads to the construction of an energy model representing the operating modes for each device, as well as the total amount of energy in the system. This model includes all the influential hardware/software energy constraints, derived from the simulation in the previous step. Those constraints are added in the energy model in the form of probabilistic distributions using a distribution fitting technique similar to the one presented in \cite{lekidis2015design}.
\item{\bf Calibration for the construction of an energy-aware System Model}: This step concerns with the addition of parameters for the runtime characterization of the IoT application with respect to the BIP System Model as well as the generation of glueing code for the combination of the BIP System Model with the energy model obtained from the previous steps. The combination leads to the construction of the Calibrated BIP System Model, which allows to analyze energy aspects and to evaluate energy requirements. 
\item{{\bf Requirement formalization:} This step concerns with the process of expressing a requirement with temporal logic properties. The properties are derived from the input requirement specification, where they are expressed in natural language. The resulting properties of this step are used as input in step 7. }
\item{{\bf Statistical model checking (SMC):} In order to verify the model against the requirements we use SMC. The resulting verification verdict allows to find tight bounds for the energy consumed in the IoT application as well as in the individual devices. }
\end{enumerate}

\subsection{Energy-aware parameter configuration} \label{sec:energyParameters}

As a first step of the proposed method, the XML-based energy-aware parameter configuration is translated into a CSC file (Step 1 of Figure \ref{fig:flow}). To better understand the importance of the selected parameters for the model we hereby introduce the three categories of parameters that affect the overall energy consumption, namely (i) the application layer, (ii) the MAC layer and (ii) the physical layer parameters.
Each category introduces parameters that depend on each other, but there are no inter-dependencies between the different categories. For instance, the choice of the IoT application protocol in the application layer is independent from the choice of the duty-cycling mechanisms in the MAC layer.

\noindent\textbf{MAC layer}: The Contiki network stack uses a Radio Duty Cycling (RDC) mechanism, allowing a device radio to remain active and listen only for certain periods in time. The remaining time it moves in the LPM mode, where it cannot receive any transmitted  packet. This mechanism reduces the consumed energy, as the radio remains in listening mode for significantly less time. RDC mechanisms are divided into 1) synchronous and 2) asynchronous, according to the technique that is used for awakening the IoT device. The former allow to maintain a Time-Division Multiple Access (TDMA) mechanism period, with which the device wakes up only in the beginning of the period to receive any incoming packets for a fixed time interval and afterwards switches to LPM mode. The latter allow a device to receive only the packets destined for it. This is possible through a Sender-initiated mode, where the sender transmits frequently a preamble packet, that is acknowledged only when the Receiver is awake, or a Receiver-initiated, where the Receiver transmits a broadcast Probe Packet to all IoT network devices to indicate that it is awake. Contiki implements various mechanisms for RDC that are Sender- or Receiver-initiated and have different characteristics. These characteristics are considered as the main parameters that influence the energy consumption in RDC and are:

\begin{enumerate}

\item{\textit{RDC protocol}}: This parameter refers to the protocol that is used on top of the MAC protocol in the Contiki network stack, to implement the duty cycling for the IoT devices. The selection of this protocol affects significantly the overall consumed energy, as it allows the device to switch between the operating modes with different mechanisms. Therefore, the choice is also present as one of the parameters of the XML-based energy configuration. In Contiki, there are four allowed values for this choice, namely the ContikiMAC, the X-MAC, the Low Power Probing (LPP) \cite{eriksson2009cooja} as well as the no use of RDC protocol named as nullRDC. ContikiMAC is a protocol based on the principles behind low-power listening but with better power efficiency. X-MAC is based on the original X-MAC protocol \cite{buettner2006x}, but has been enhanced to reduce power consumption and maintain good network conditions. Contiki's LPP is based on asynchronous receiver-initiated transmission scheme protocol, where the devices wake up and send a probe to inform other devices on their receiving availability. 

\item{\textit{RDC frequency}}: This parameter specifies the frequency with which the IoT devices wake-up to check the channel for any packets whose transmission is pending through their RDC mechanism. Channel checking is accomplished by transmitting probe packets to verify any existing activity. In the presence of channel activity they remain in the Rx operating mode to receive any transmitted packets from the other IoT network devices, otherwise they switch to the LPM mode for another duty-cycling period. By remaining longer in Rx mode an IoT device has increased energy consumption, therefore we consider the RDC frequency as a significant parameter in the energy parameter XML configuration. 

\item{\textit{Packet retransmissions}}: IoT applications are often prone to transmission errors in the MAC layer that are leading to successive loss of packets, which are usually caused by extensive loss of network bandwidth. For this reason wireless protocols for IoT usually include a mechanism handling the retransmission of non-acknowledged packets in the MAC layer. This mechanism increases the IoT system reliability, but also keeps the IoT device in the Tx operating mode for longer time durations. Therefore, the overall energy consumed by an IoT device is increased, making this parameter part of the energy parameter configuration in our design method. 
\end{enumerate}

\noindent\textbf{Application layer}: A Contiki IoT application can use a variety of protocols for data communication. Each protocol offers different mechanisms and is characterized by different energy consumption. It is also possible to introduce a different size for the packets to be transmitted. For this reason we define the application layer parameters as follows: 

\begin{enumerate}
\setcounter{enumi}{3}
\item{\textit{Application protocol}: This parameter refers to the protocol that is chosen for data exchange. This choice depends strongly on the application requirements i.e. safety critical applications require reliable data transfer through MQTT communication, whereas wireless sensor applications require energy efficiency and faster communications, therefore they rather rely on CoAP communication. Nevertheless, since the protocols offer different mechanisms, this choice has a significant impact on the lifetime of an IoT device.}
\item{\textit{Header size}: IoT applications usually include compression algorithms to reduce the packet header, in order to reduce the overall size of the exchanged message \cite{lekidis2015design}. Such algorithms ensure the minimal radio usage period for data transmission, therefore minimizing the cumulative energy consumption. However, in IoT devices this is a trade-off since for the computation time of compression/decompression of the packets, the IoT device remains in the CPU operating mode. For these reasons, the header size is considered as a parameter in our input configuration file. }   

%

\end{enumerate}
\noindent\textbf{Communication medium}: IoT applications usually contain wireless devices that communicate over a communication medium which is prone to errors (e.g. packet collisions) that impact the energy consumption. This category includes the following parameters: 

\begin{enumerate}
\setcounter{enumi}{5}
\item{\textit{Radio interference}: This parameter is related to the presence of interference in the communication medium as a form of additive noise from simultaneous transmissions of proximity networks with the same radio frequency. Interference leads to increased packet collisions that can impact the energy consumption in the IoT devices, as they remain in the Tx operating mode for longer time durations. This parameter is added to the energy parameter configuration by introducing and properly configuring the disturber mote type as a part of the Contiki IoT application \cite{boano2009controllable}.}
\end{enumerate}

When the previous analysis is expanded to the entire IoT system architecture we ought to find additional parameters impacting the energy consumption. Such a characteristic parameter is the routing protocol that is used along with IPv4/IPv6 in the networking layer of the Contiki protocol stack to ensure the connection of Wide Area Network (WAN) IoT networks (e.g. meshed or multi-hop). In this scenario IoT devices have to be configured as a border router \cite{kovatsch2011low} to connect distant networks, which requires the extensive use of communication bandwidth. The use of the Routing Protocol for Low power and Lossy Networks (RPL) routing mechanisms increases computation/communication time durations in the IoT devices as well as in the overall system.         

\subsection{Energy model} \label{sec:energyModel}

As a part of step 4 of the design flow, an energy model is derived reflecting the true energy consumption in the BIP System Model. The main equations used to compute energy constraints in the Contiki IoT environment are as follows. Initially, the duty cycle (\textit{D}) as a device ratio is computed by: 
\begin{equation} \label{Eq:dutyCycle}
D_y = \dfrac{\mathop{\sum_{i=1}^{N_y} I_{y}*V_{y}*\Delta t_{{y}_{i}}}}{E_{total}}
\end{equation}

\noindent 
where y indicates the operating mode for the IoT device and $N_y$ the relative number of occurrences that the device visits the respective operating mode y, given that it cannot be in two modes within the same time interval. Based on the duty cycle of a device in a given operating mode, we can also derive the overall energy consumption (in Joule) over every device operating mode:

\begin{equation} 
\begin{aligned} \label{eq:totalEnergy}
E_{total}= \mathop{\sum_{\forall i \in D_{LPM}}^{N_{LPM}} I_{LPM}*V_{LPM}*\Delta t_{{LPM}_{i}}} +  \mathop{\sum_{\forall j \in D_{Tx}}^{N_{Tx}}I_{Tx}*V_{Tx}*\Delta t_{{Tx}_{j}}} \\ + \mathop{\sum_{\forall k \in D_{Rx}}^{N_{Rx}}I_{Rx}*V_{Rx}*\Delta t_{{Rx}_{k}}}+\mathop{\sum_{\forall z \in D_{CPU}}^{N_{CPU}}I_{CPU}*V_{CPU}*\Delta t_{{CPU}_{z}}}+ E_{PER}
\end{aligned}
\end{equation}

\noindent
where in every tuple $\mathop{\{Iy,Vy,\Delta t{_y}\}}$  y indicates the operating mode as in Equation \ref{Eq:dutyCycle} and $I,V,$ and $\Delta$ indicate respectively the current (in Ampere), voltage (in Volts) and time intervals in which the device remains in each operating mode. The sum is over the number of occurrences $N_{LPM},N_{Tx},N_{Rx},N_{CPU}$ of the device visiting the respective operating mode y and $D_y$ indicates the duty cycle for each mode. $E_{PER}$ indicates the energy consumed by the device peripherals (in Joule). Finally, the device lifetime (\textit{lf}) is computed by: 
\begin{equation}
lf=\dfrac{C_{batt}* Vcc}{E_{total}} 
\end{equation}

\noindent 
where $C_{batt}$ indicates the overall capacity of the battery for autonomous operation (in Ampere hours), Vcc the operating voltage (in Volts) and $E_{total}$ the total amount of energy. 

\begin{figure}
\begin{minipage}[bthp!]{1\textwidth}
\centering
\includegraphics[width=0.55\textwidth]{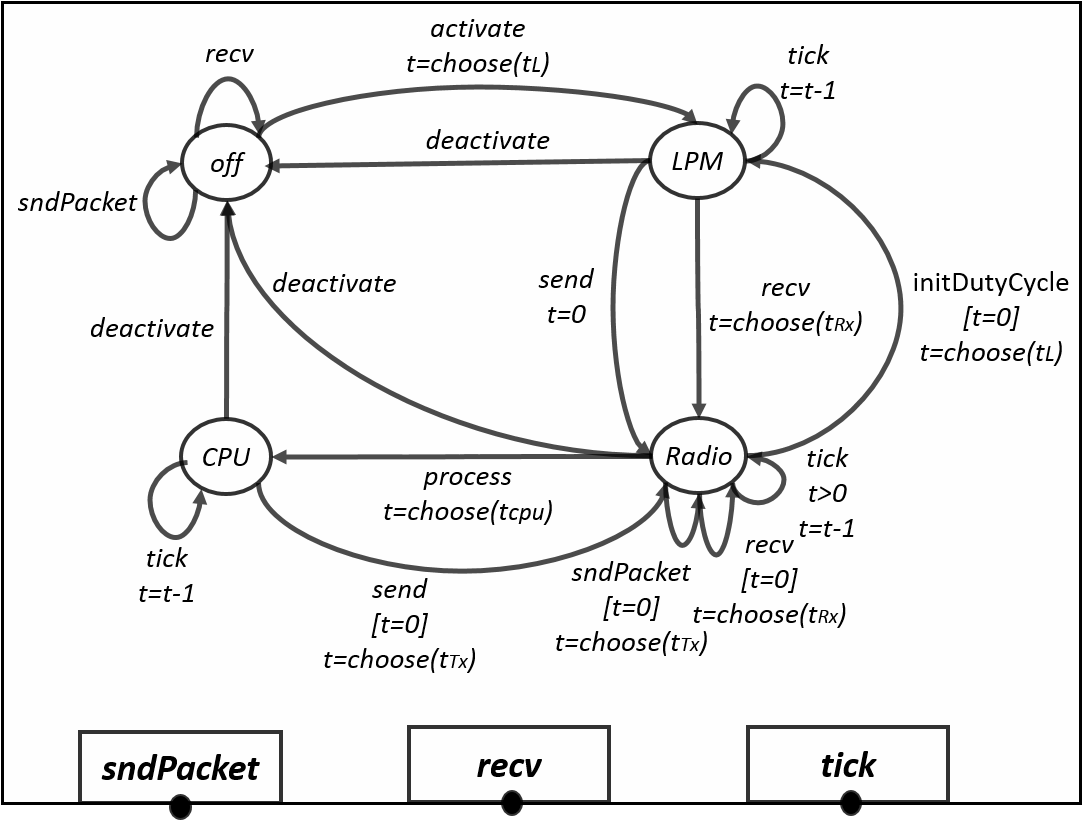}
\caption{BIP energy model}
\label{fig:energyModel}
\end{minipage}
\end{figure}

In Figure \ref{fig:energyModel} we illustrate a fragment of the BIP energy model that is derived in Step 4 of our method. We also present in Listing \ref{Code-1} the code with associated actions as in the model to facilitate its description.


By using the $powertrace\_start$ command in Line 11 of Listing \ref{Code-1} the radio is activated and it immediately switches to the LPM operating mode. This is illustrated in Figure \ref{fig:energyModel} through the $activate$ transition. In line 16 the CoAP message is initiated, which triggers the $sndPacket$ transition in the model. Before a packet is transmitted the device has to initialize its header and payload as its shown in lines 17-19. This is represented with the $process$ transition in the model, which switches it to the CPU state (i.e. indicating the CPU processing mode). When processing is finished the packet is transmitted through the $sndPacket$ transition. If no further transmissions or receptions (through the transition $recv$) of packets take place, the model will return to the LPM state for another duty-cycle through the $initDutyCycle$ transition. The durations considered in every state or transition of the model are selected from probabilistic distributions that are gathered from profiling the generated code that is simulated in Cooja \cite{lekidis2015design}. The gathered distributions correspond to each one of the powertrace operating modes. Finally, the model contains 3 exported transitions (i.e. $sndPacket$, $recv$, $tick$) to interact with the OS/Kernel components of the BIP system model \cite{lekidis2018model} that is constructed in Step 2 of our method. 

\lstset{language=C,
		captionpos=b,
		numbers=left,
		basicstyle=\ttfamily\scriptsize,
		numbersep=8pt,
		frame=single,
		framexleftmargin=15pt,
		framexrightmargin=15pt,
		showstringspaces=false,
} 

\begin{figure}[bthp]
\centering
\begin{minipage}[b]{0.7\textwidth}
\begin{lstlisting}[caption={Contiki powertrace client}, label={Code-1}] 
#include "contiki.h" 
#include "powertrace.h" 
PROCESS(power, "powertrace example"); 
AUTOSTART_PROCESSES(&power); 
PROCESS_THREAD(power, ev, data) 
{ 
  static struct etimer et; 
  PROCESS_BEGIN(); 
  /* Start powertracing */ 
  int RTIMER = 1; // 1 second reporting cycle 
  powertrace_start(CLOCK_SECOND * RTIMER); 
  etimer_set(&et, CLOCK_SECOND*t); 
  while(1) {  
    ....
    if(etimer_expired(&et)) {
      coap_init_message(request, COAP_TYPE_CON, COAP_POST, 0);
      coap_set_payload(request, (uint8_t *)msg, sizeof(msg) - 1);
      COAP_BLOCKING_REQUEST(&server_ipaddr, REMOTE_PORT, 
      	request, client_chunk_handler);
  }
} 
PROCESS_END(); 

\end{lstlisting} 
\end{minipage}
\end{figure}

\section{Case-study: Energy-aware building management system}\label{sec:caseStudy}
In this section we present the case-study for validating our method in the context of a Building Management System (BMS). The application aims at automating the operation of the basic structures in a building installation connected through a dedicated WPAN network for each building floor. Additionally, it also allows for the remote control of different buildings through a WAN network. Such an architectural setup is employed by new-generation building facilities \footnote{\url{https://www.intel.com/content/www/us/en/smart-buildings/overview.html}}. In particular, the case-study describes a company building, consisting of different floors with office rooms to which workers have access during working-hours (8.00 till 18.00).

\begin{figure}[htbp!]
\begin{centering}
\includegraphics[width=1\textwidth]{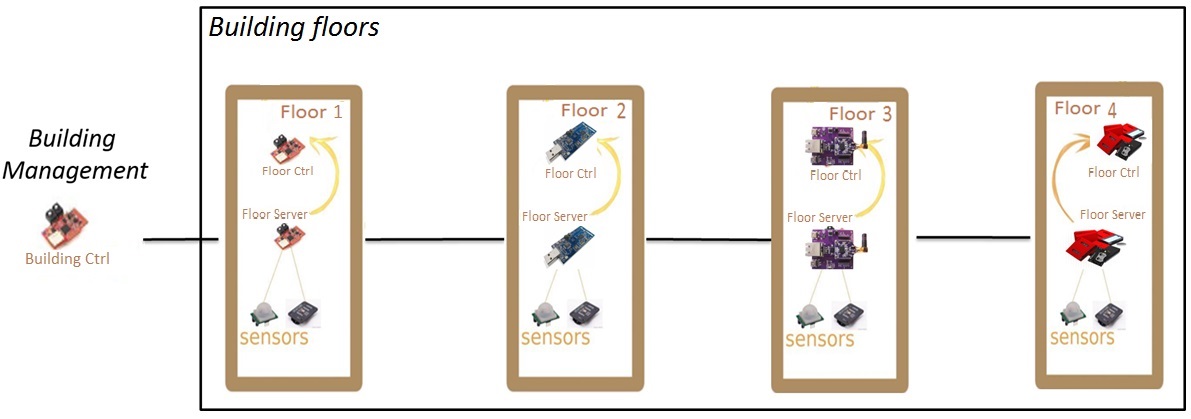}
\caption{Architectural topology in the building management system}
\label{fig:caseStudyOverview}
\end{centering}
\end{figure}

Figure \ref{fig:caseStudyOverview} presents the heterogeneous IoT system architecture, where each floor has two unique IoT devices in the roles of a floor controller and floor server. The architectural heterogeneity allows to analyze energy consumption in a variety of IoT devices, such as Zolertia Z1 (level 1), Sky mote (level 2), OpenMote (level 3) and SimpleLink Sensortag (level 4). 

Each set of floor devices communicates its state to the lower-level and the lowest level (floor 1 in Figure \ref{fig:caseStudyOverview}) forwards all the floor states to the Building Management device located in level 1. This device forms the central system supervisor. It is notified by all floor controllers for their current status and can be accessed or managed remotely as part of the WAN network. The state of each floor is determined by a set of sensor/actuator resources present in the floor server devices. In this case-study, we used the common resources amongst the four considered device types i.e. temperature sensor, humidity sensor, motion sensor, light sensor/actuator, alarm actuator, light actuator, thermostat actuator.

Concerning the system functionality, each floor server is a node in which every sensor/actuator is represented as a REST endpoint through which the floor controller can at any time know the current state of the floor. For example, for the temperature sensor the floor controller monitors the temperature in the floor and in case it exceeds the user-defined upper or lower limits, then it switches on the thermostat. Additionally, the building also employs an energy-saving mechanism, in order to automatically diminish the thermostat limits during non-working hours. Moreover,
the floor controller is also subscribed to the motion sensor, as well as to the light sensor/actuator, in order to detect motion and open the lights if the ambient light in the environment is not sufficient. The corresponding lights that open as a result of motion detection are linked to the location, where motion was detected.

\subsection{Application of the proposed method} \label{sec:methodApplication}
We focus now on the individual steps of the design flow in Figure \ref{fig:flow} for the outlined BMS application.
\noindent \domanda{\textnormal{\textit{Step 1: Translation  of the energy parameter configuration}}}
The parameters in Section \ref{sec:energyParameters} that influence the energy consumption obtain their actual values based on the application requirements. For the BMS application, we were based on the default values taken from the kernel libraries of the Contiki OS (Table \ref{tab:modelParam}). Moreover, to better demonstrate the impact of the parameters, we added as a part of our simulations a variation range. Table \ref{tab:modelParam} introduces this variation range along with the default values in the Contiki OS. 

\begin{table}[htbp]
  \centering
    \resizebox{12.2cm}{!} {
    \begin{tabular} {|c||c||c|}
        \hline
    \textbf{Energy model parameter} & \textbf{Default value} & \textbf{Variation range} \\
    \hline
          RDC protocol & X-MAC & [Contiki-MAC, X-MAC, LPP, nullRDC] \\
          RDC frequency & 8 Hz & [2-32] Hz (even number) \\
          Packet retransmissions  & 4 & [0-5] \begin{math} \in \mathbb{Z} \end{math}  \\
          Service protocol & CoAP & [CoAP, MQTT, HTTP] \\
          Header size & 48 bytes & [32-64] bytes (even number) \\
          Interference & 0 & [0-1] \begin{math} \in \mathbb{R} \end{math} \\
    \hline
    \end{tabular}}%
  \caption{Energy parameters of the XML configuration}
  \label{tab:modelParam}%
\end{table}%
\noindent Accordingly, in \textit{Steps 2} (\textit{Transformation for the System Model}) and \textit{3} (\textit{Code generation}) we used the described XML configuration along with the DSL to generate the Contiki Simulation Configuration as well as the BIP System Model. 
\vspace{-.5cm}
\noindent \domanda{\textnormal{\textit{Step 4: Energy characterization}}}
\nobreak\noindent\ignorespaces Due to the device heterogeneity for each floor in the BMS application, we could only apply the distribution fitting technique to energy data coming from the same device type. We then fitted the data into Poisson distributions for Tx and CPU modes, whereas the energy for Rx and LPM mode followed a normal distribution during the course of a day. The resulting distributions of this step are used in \textit{Step 5} (\textit{Calibration for the construction of an energy-aware System Model}) to calibrate the model of Figure \ref{fig:energyModel}.  
\vspace{-.5cm}
\noindent \domanda{\textnormal{\textit{Step 6: Formalization of system requirements}}}
We identified three requirements for the BMS system, from which only the 
first one concerns the IoT device lifetime, whereas 
the remaining two concern the IoT device duty-cycle in different operating modes. These 
requirements are:

\noindent{\bf Requirement 1. }Device lifetime should be at least 1 week. \\
\noindent{\bf Requirement 2. }The duty-cycle in the LPM mode should remain higher than 90\% during working hours. \\
\noindent{\bf Requirement 3. }The duty-cycle in the Rx mode should not exceed 20\% during working hours.

\subsection{Experiments}

In this section we demonstrate the experiments for evaluating the aforementioned requirements. The current and voltage values for each operating mode that were used for the calculation of the total energy, duty cycle and device lifetime in the Section \ref{sec:energyModel} equations were obtained from the IoT devices' datasheet.

We use the parameters of Section \ref{sec:energyParameters} to demonstrate the impact of a certain parameter to the overall energy consumed in a device. To this end, we focused on a certain category parameter and experimented with all variations of the other parameters in the same category, while the parameters in different categories were set in their default values (Table \ref{tab:modelParam}). This is due to the independence between the categories. 


The system requirements were validated through the continuous simulation of the BIP system model for a working week, where we used SMC to measure the probability for satisfying the requirements of Section \ref{sec:methodApplication}.

\begin{figure}[htbp!]
\begin{centering}
\includegraphics[width=1\textwidth]{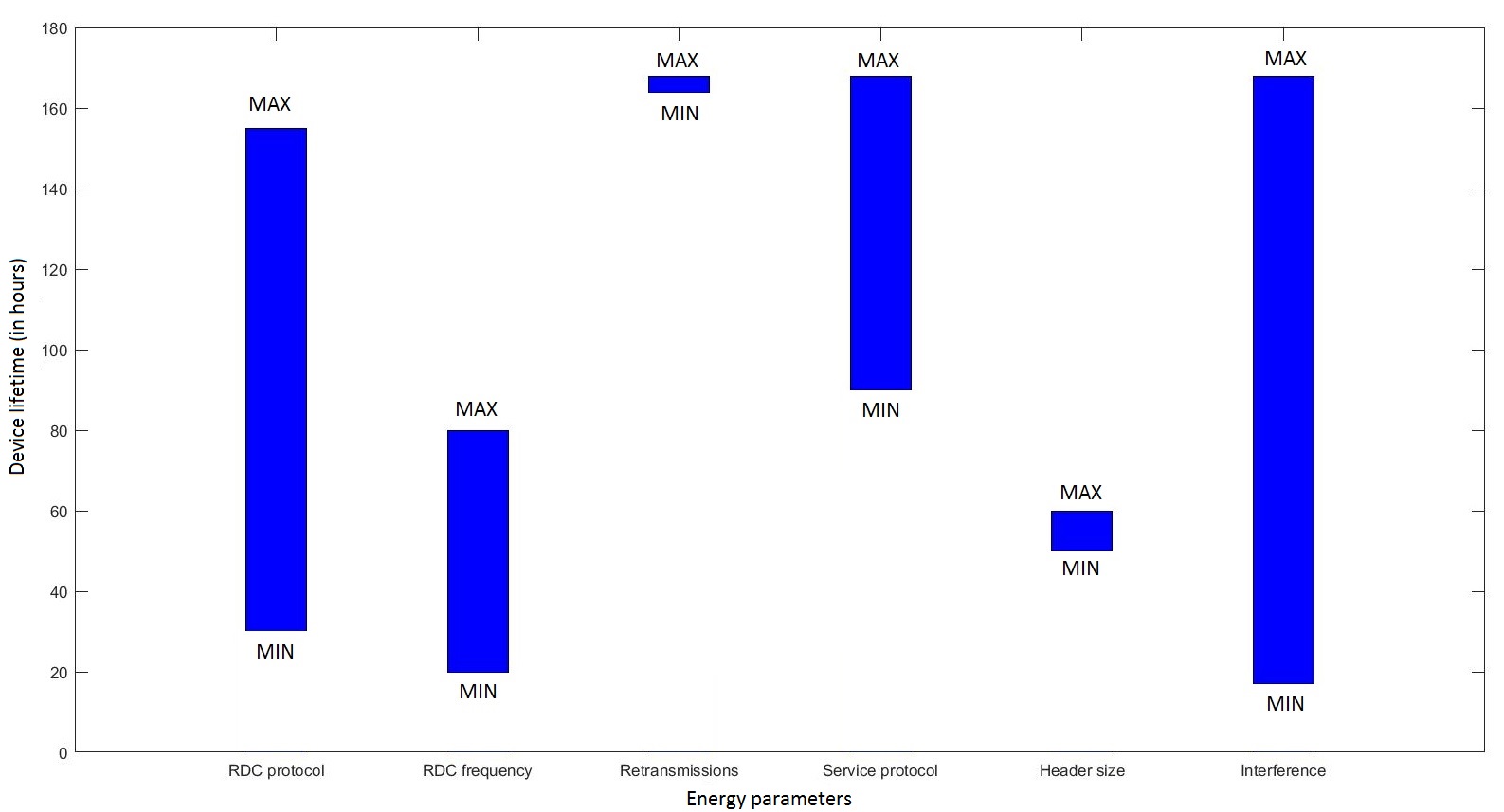}
\caption{Device lifetime for different variations of energy parameters}
\label{fig:lifetime}
\end{centering}
\end{figure}

\noindent{\bf Requirement 1.} We evaluated the property $\phi_1 = lf\geq 168$, where 168 indicates the sum of hours during the week. For this experiment we used two scenarios the first being with the default values of Table \ref{tab:modelParam} and the second with different sets of the parameters of the same table. For the first scenario we found that $P(\phi_1) = 0.9$, whereas for the second scenario we have conducted several sets of experiments. These experiments allowed us to demonstrate the contribution of the energy parameters to the device lifetime (Figure \ref{fig:lifetime}). The greatest lifetime impact is observed in experiments involving the RDC protocol parameter as the difference between maximum and minimum battery autonomy is 130 hours, whereas with radio interference the difference is 148. On the other hand, we also observe that the variation in the retransmissions parameter has no impact on the system, as the BMS network of our case-study is formed by several smaller WPAN networks with a small number of IoT devices. Furthermore, this Figure allows the system designer to know the values to be used for the parameters of Table \ref{tab:modelParam} to satisfy this requirement.

\begin{figure}[htbp!]
\begin{centering}
\includegraphics[width=0.9\textwidth]{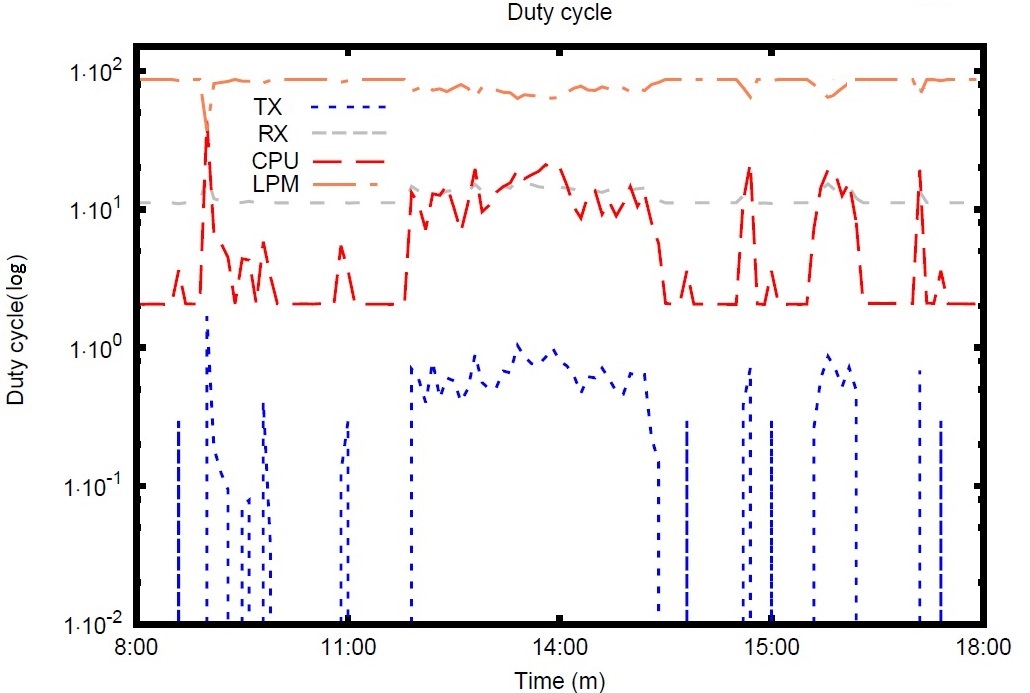}
\caption{Duty cycle of the Zolertia floor controller during a working day using the LPP RDC protocol (in logarithmic scale)} 
\label{fig:duty-cycle}
\end{centering}
\end{figure}

\noindent{\bf Requirement 2.} We verified the property $\phi_2 = D_{LPM}\geq 90\%$. The property is significantly influenced by the RDC protocol parameter. In particular, it holds for the LPP RDC protocol ($P(\phi_2) = 1$) as it is illustrated by Figure \ref{fig:duty-cycle}, which focuses on the duty cycle of each operating mode for the specific RDC protocol. In the same Figure, since the energy in certain modes (i.e LPM) during our experiments was significantly higher than in all the remaining operating modes we chose to present our results in logarithmic scale. The reasoning behind the energy saved with LPP in LPM is that the sum of packets is exchanged during small and continuous time intervals, while in all the remaining time intervals the device switches to a deep sleep mode. Concerning the other RDC protocols for ContikiMAC and XMAC, we found $P(\phi_2) = 0.5$ and $P(\phi_2) = 0.7$ respectively. The difference between the two RDC protocols is explained by the packet transmission in ContikiMAC rather than a pulse as in X-MAC, in order to activate the wake-up signal. Finally, for the nullRDC protocol $P(\phi_2) = 0$ as the radio remains in Rx for long time intervals.

\noindent{\bf Requirement 3.} We verified the property $\phi_3 = D_{Rx}\leq 20\%$. The property holds when the chosen RDC protocol is LPP (Figure \ref{fig:duty-cycle}). However, with the two other protocols, the observed probability was $P(\phi_3) = 0.8$ for XMAC, $P(\phi_3) = 0.6$ for ContikiMAC and $P(\phi_3) = 0$ for the nullRDC. This is due to the different awakening frequency of the device for listening incoming packets, especially during working hours when the data exchange is intense (Figure \ref{fig:duty-cycle}). 

\section{Conclusion} \label{sec:conc}
We presented a novel method for profiling and characterizing the energy consumption in IoT applications, as well as in the individual IoT devices. The method is based on the principles of rigorous system design by using the BIP component framework. Currently, the provided support concerns with the design of REST service-based applications which are deployed on nodes running the Contiki OS. The method takes as input the application design description in DSL and an XML-based set of energy parameters and generates a system model in BIP for validating requirements related to energy characteristics. The system model is calibrated with energy constraints that are obtained by the simulation of the code generated from the application description. The calibrated model is afterwards used to validate the requirements through Statistical Model Checking (SMC). 

As a proof of concept, the described method has been applied to a building management system. The system consists of several subsystems deployed in multiple floors of a smart building facility using several well-known IoT devices. We have verified the energy-related requirements concerning the device lifetime, as well as the duty-cycle of the devices and the overall system. The results allow us to characterize the energy consumption on the system and calculate accurately the device lifetime.

As future work, it is worth to consider the aspect of remote control in the building, as well as its impact on the overall energy consumption. This can be accomplished through the presence of border routers and RPL routing mechanisms in the IoT application. In this direction, an important architectural characteristic of IoT systems is fog and cloud computing \cite{bonomi2012fog}, where a significant part of the computation is no longer handled by the resource-constrained IoT devices. Hence, the overall energy consumption in the system is reduced. Additionally, the Building Management Controller that was presented in the case-study can also perform control actions based on the data it gathers, such as shutting down the heating and lighting system if there is no motion for certain hours during the day.

\bibliographystyle{eptcs}
\bibliography{generic}

\begin{thebibliography}{10}
\providecommand{\bibitemdeclare}[2]{}
\providecommand{\surnamestart}{}
\providecommand{\surnameend}{}
\providecommand{\urlprefix}{Available at }
\providecommand{\url}[1]{\texttt{#1}}
\providecommand{\href}[2]{\texttt{#2}}
\providecommand{\urlalt}[2]{\href{#1}{#2}}
\providecommand{\doi}[1]{doi:\urlalt{http://dx.doi.org/#1}{#1}}
\providecommand{\bibinfo}[2]{#2}

\bibitemdeclare{article}{basu2011rigorous}
\bibitem{basu2011rigorous}
\bibinfo{author}{Ananda \surnamestart Basu\surnameend},
  \bibinfo{author}{Bensalem \surnamestart Bensalem\surnameend},
  \bibinfo{author}{Marius \surnamestart Bozga\surnameend},
  \bibinfo{author}{Jacques \surnamestart Combaz\surnameend},
  \bibinfo{author}{Mohamad \surnamestart Jaber\surnameend},
  \bibinfo{author}{Thanh-Hung \surnamestart Nguyen\surnameend} \&
  \bibinfo{author}{Joseph \surnamestart Sifakis\surnameend}
  (\bibinfo{year}{2011}): \emph{\bibinfo{title}{Rigorous component-based system
  design using the BIP framework}}.
\newblock {\sl \bibinfo{journal}{IEEE software}}
  \bibinfo{volume}{28}(\bibinfo{number}{3}), pp. \bibinfo{pages}{41--48},
  \doi{10.1109/MS.2011.27}.

\bibitemdeclare{inproceedings}{benini1998system}
\bibitem{benini1998system}
\bibinfo{author}{Luca \surnamestart Benini\surnameend}, \bibinfo{author}{Robin
  \surnamestart Hodgson\surnameend} \& \bibinfo{author}{Polly \surnamestart
  Siegel\surnameend} (\bibinfo{year}{1998}): \emph{\bibinfo{title}{System-level
  power estimation and optimization}}.
\newblock In: {\sl \bibinfo{booktitle}{Proceedings of the 1998 international
  symposium on Low power electronics and design}}, \bibinfo{organization}{ACM},
  pp. \bibinfo{pages}{173--178}, \doi{10.1145/280756.280881}.

\bibitemdeclare{inproceedings}{boano2009controllable}
\bibitem{boano2009controllable}
\bibinfo{author}{Carlo~Alberto \surnamestart Boano\surnameend},
  \bibinfo{author}{Zhitao \surnamestart He\surnameend}, \bibinfo{author}{Yafei
  \surnamestart Li\surnameend}, \bibinfo{author}{Thiemo \surnamestart
  Voigt\surnameend}, \bibinfo{author}{Marco \surnamestart
  Z{\'u}{\~n}niga\surnameend} \& \bibinfo{author}{Andreas \surnamestart
  Willig\surnameend} (\bibinfo{year}{2009}): \emph{\bibinfo{title}{Controllable
  radio interference for experimental and testing purposes in wireless sensor
  networks}}.
\newblock In: {\sl \bibinfo{booktitle}{Local Computer Networks, 2009. LCN 2009.
  IEEE 34th Conference on}}, \bibinfo{organization}{IEEE}, pp.
  \bibinfo{pages}{865--872}, \doi{10.1109/LCN.2009.5355013}.

\bibitemdeclare{inproceedings}{bonomi2012fog}
\bibitem{bonomi2012fog}
\bibinfo{author}{Flavio \surnamestart Bonomi\surnameend},
  \bibinfo{author}{Rodolfo \surnamestart Milito\surnameend},
  \bibinfo{author}{Jiang \surnamestart Zhu\surnameend} \&
  \bibinfo{author}{Sateesh \surnamestart Addepalli\surnameend}
  (\bibinfo{year}{2012}): \emph{\bibinfo{title}{Fog computing and its role in
  the internet of things}}.
\newblock In: {\sl \bibinfo{booktitle}{Proceedings of the first edition of the
  MCC workshop on Mobile cloud computing}}, \bibinfo{organization}{ACM}, pp.
  \bibinfo{pages}{13--16}, \doi{10.1145/2342509.2342513}.

\bibitemdeclare{inproceedings}{buettner2006x}
\bibitem{buettner2006x}
\bibinfo{author}{Michael \surnamestart Buettner\surnameend},
  \bibinfo{author}{Gary~V \surnamestart Yee\surnameend}, \bibinfo{author}{Eric
  \surnamestart Anderson\surnameend} \& \bibinfo{author}{Richard \surnamestart
  Han\surnameend} (\bibinfo{year}{2006}): \emph{\bibinfo{title}{X-MAC: a short
  preamble MAC protocol for duty-cycled wireless sensor networks}}.
\newblock In: {\sl \bibinfo{booktitle}{Proceedings of the 4th international
  conference on Embedded networked sensor systems}},
  \bibinfo{organization}{ACM}, pp. \bibinfo{pages}{307--320},
  \doi{10.1145/1182807.1182838}.

\bibitemdeclare{inproceedings}{colitti2011rest}
\bibitem{colitti2011rest}
\bibinfo{author}{Walter \surnamestart Colitti\surnameend},
  \bibinfo{author}{Kris \surnamestart Steenhaut\surnameend},
  \bibinfo{author}{Niccolo \surnamestart De~Caro\surnameend},
  \bibinfo{author}{Bogdan \surnamestart Buta\surnameend} \&
  \bibinfo{author}{Virgil \surnamestart Dobrota\surnameend}
  (\bibinfo{year}{2011}): \emph{\bibinfo{title}{REST enabled wireless sensor
  networks for seamless integration with web applications}}.
\newblock In: {\sl \bibinfo{booktitle}{Mobile Adhoc and Sensor Systems (MASS),
  2011 IEEE 8th International Conference on}}, \bibinfo{organization}{IEEE},
  pp. \bibinfo{pages}{867--872}, \doi{10.1109/MASS.2011.102}.

\bibitemdeclare{inproceedings}{dunkels2007software}
\bibitem{dunkels2007software}
\bibinfo{author}{Adam \surnamestart Dunkels\surnameend},
  \bibinfo{author}{Fredrik \surnamestart Osterlind\surnameend},
  \bibinfo{author}{Nicolas \surnamestart Tsiftes\surnameend} \&
  \bibinfo{author}{Zhitao \surnamestart He\surnameend} (\bibinfo{year}{2007}):
  \emph{\bibinfo{title}{Software-based on-line energy estimation for sensor
  nodes}}.
\newblock In: {\sl \bibinfo{booktitle}{Proceedings of the 4th workshop on
  Embedded networked sensors}}, \bibinfo{organization}{ACM}, pp.
  \bibinfo{pages}{28--32}, \doi{10.1145/1278972.1278979}.

\bibitemdeclare{inproceedings}{eriksson2009cooja}
\bibitem{eriksson2009cooja}
\bibinfo{author}{Joakim \surnamestart Eriksson\surnameend},
  \bibinfo{author}{Fredrik \surnamestart {\"O}sterlind\surnameend},
  \bibinfo{author}{Niclas \surnamestart Finne\surnameend},
  \bibinfo{author}{Nicolas \surnamestart Tsiftes\surnameend},
  \bibinfo{author}{Adam \surnamestart Dunkels\surnameend},
  \bibinfo{author}{Thiemo \surnamestart Voigt\surnameend},
  \bibinfo{author}{Robert \surnamestart Sauter\surnameend} \&
  \bibinfo{author}{Pedro~Jos{\'e} \surnamestart Marr{\'o}n\surnameend}
  (\bibinfo{year}{2009}): \emph{\bibinfo{title}{COOJA/MSPSim: interoperability
  testing for wireless sensor networks}}.
\newblock In: {\sl \bibinfo{booktitle}{Proceedings of the 2nd International
  Conference on Simulation Tools and Techniques}}, \bibinfo{organization}{ICST
  (Institute for Computer Sciences, Social-Informatics and Telecommunications
  Engineering)}, p.~\bibinfo{pages}{27}, \doi{10.4108/ICST.SIMUTOOLS2009.5637}.

\bibitemdeclare{article}{georgiou2017iot}
\bibitem{georgiou2017iot}
\bibinfo{author}{Kyriakos \surnamestart Georgiou\surnameend},
  \bibinfo{author}{Samuel \surnamestart Xavier-de Souza\surnameend} \&
  \bibinfo{author}{Kerstin \surnamestart Eder\surnameend}
  (\bibinfo{year}{2017}): \emph{\bibinfo{title}{The IoT energy challenge: A
  software perspective}}.
\newblock {\sl \bibinfo{journal}{IEEE Embedded Systems Letters}},
  \doi{10.1109/LES.2017.2741419}.

\bibitemdeclare{inproceedings}{herault2004approximate}
\bibitem{herault2004approximate}
\bibinfo{author}{T.~\surnamestart H{\'e}rault\surnameend},
  \bibinfo{author}{R.~\surnamestart Lassaigne\surnameend},
  \bibinfo{author}{F.~\surnamestart Magniette\surnameend} \&
  \bibinfo{author}{S.~\surnamestart Peyronnet\surnameend}
  (\bibinfo{year}{2004}): \emph{\bibinfo{title}{Approximate probabilistic model
  checking}}.
\newblock In: {\sl \bibinfo{booktitle}{Verification, Model Checking, and
  Abstract Interpretation}}, \bibinfo{organization}{Springer}, pp.
  \bibinfo{pages}{73--84}, \doi{10.1007/978-3-540-24622-0_8}.

\bibitemdeclare{article}{hoeffding1963probability}
\bibitem{hoeffding1963probability}
\bibinfo{author}{W.~\surnamestart Hoeffding\surnameend} (\bibinfo{year}{1963}):
  \emph{\bibinfo{title}{Probability inequalities for sums of bounded random
  variables}}.
\newblock {\sl \bibinfo{journal}{Journal of the American statistical
  association}} \bibinfo{volume}{58}(\bibinfo{number}{301}), pp.
  \bibinfo{pages}{13--30}, \doi{10.1214/aoms/1177730491}.

\bibitemdeclare{inproceedings}{kovatsch2011low}
\bibitem{kovatsch2011low}
\bibinfo{author}{Matthias \surnamestart Kovatsch\surnameend},
  \bibinfo{author}{Simon \surnamestart Duquennoy\surnameend} \&
  \bibinfo{author}{Adam \surnamestart Dunkels\surnameend}
  (\bibinfo{year}{2011}): \emph{\bibinfo{title}{A low-power CoAP for Contiki}}.
\newblock In: {\sl \bibinfo{booktitle}{Mobile Adhoc and Sensor Systems (MASS),
  2011 IEEE 8th International Conference on}}, \bibinfo{organization}{IEEE},
  pp. \bibinfo{pages}{855--860}, \doi{10.1109/MASS.2011.100}.

\bibitemdeclare{inproceedings}{legay2010statistical}
\bibitem{legay2010statistical}
\bibinfo{author}{Axel \surnamestart Legay\surnameend},
  \bibinfo{author}{Beno{\^\i}t \surnamestart Delahaye\surnameend} \&
  \bibinfo{author}{Saddek \surnamestart Bensalem\surnameend}
  (\bibinfo{year}{2010}): \emph{\bibinfo{title}{{Statistical model checking: An
  overview}}}.
\newblock In: {\sl \bibinfo{booktitle}{Runtime Verification}},
  \bibinfo{organization}{Springer}, pp. \bibinfo{pages}{122--135},
  \doi{10.1016/j.ic.2006.05.002}.

\bibitemdeclare{phdthesis}{lekidis2015design}
\bibitem{lekidis2015design}
\bibinfo{author}{Alexios \surnamestart Lekidis\surnameend}
  (\bibinfo{year}{2015}): \emph{\bibinfo{title}{Design flow for the rigorous
  development of networked embedded systems}}.
\newblock Ph.D. thesis, \bibinfo{school}{Universit{\'e} Grenoble Alpes},
  \doi{10.13140/RG.2.2.19387.11042}.

\bibitemdeclare{article}{lekidis2018model}
\bibitem{lekidis2018model}
\bibinfo{author}{Alexios \surnamestart Lekidis\surnameend},
  \bibinfo{author}{Emmanouela \surnamestart Stachtiari\surnameend},
  \bibinfo{author}{Panagiotis \surnamestart Katsaros\surnameend},
  \bibinfo{author}{Marius \surnamestart Bozga\surnameend} \&
  \bibinfo{author}{Christos~K \surnamestart Georgiadis\surnameend}
  (\bibinfo{year}{2018}): \emph{\bibinfo{title}{Model-based Design of IoT
  Systems with the BIP Component Framework}}.
\newblock {\sl \bibinfo{journal}{Software – Practice and Experience}},
  \doi{10.1002/spe.2568}.

\bibitemdeclare{article}{martinez2015power}
\bibitem{martinez2015power}
\bibinfo{author}{Borja \surnamestart Martinez\surnameend},
  \bibinfo{author}{Marius \surnamestart Monton\surnameend},
  \bibinfo{author}{Ignasi \surnamestart Vilajosana\surnameend} \&
  \bibinfo{author}{Joan~Daniel \surnamestart Prades\surnameend}
  (\bibinfo{year}{2015}): \emph{\bibinfo{title}{The power of models: Modeling
  power consumption for IoT devices}}.
\newblock {\sl \bibinfo{journal}{IEEE Sensors Journal}}
  \bibinfo{volume}{15}(\bibinfo{number}{10}), pp. \bibinfo{pages}{5777--5789},
  \doi{10.1109/JSEN.2015.2445094}.

\bibitemdeclare{article}{nouri2015statistical}
\bibitem{nouri2015statistical}
\bibinfo{author}{Ayoub \surnamestart Nouri\surnameend}, \bibinfo{author}{Saddek
  \surnamestart Bensalem\surnameend}, \bibinfo{author}{Marius \surnamestart
  Bozga\surnameend}, \bibinfo{author}{Benoit \surnamestart
  Delahaye\surnameend}, \bibinfo{author}{Cyrille \surnamestart
  Jegourel\surnameend} \& \bibinfo{author}{Axel \surnamestart Legay\surnameend}
  (\bibinfo{year}{2015}): \emph{\bibinfo{title}{Statistical model checking QoS
  properties of systems with SBIP}}.
\newblock {\sl \bibinfo{journal}{International Journal on Software Tools for
  Technology Transfer}} \bibinfo{volume}{17}(\bibinfo{number}{2}), pp.
  \bibinfo{pages}{171--185}, \doi{10.1007/s10009-014-0313-6}.

\bibitemdeclare{article}{vilajosana2014realistic}
\bibitem{vilajosana2014realistic}
\bibinfo{author}{Xavier \surnamestart Vilajosana\surnameend},
  \bibinfo{author}{Qin \surnamestart Wang\surnameend}, \bibinfo{author}{Fabien
  \surnamestart Chraim\surnameend}, \bibinfo{author}{Thomas \surnamestart
  Watteyne\surnameend}, \bibinfo{author}{Tengfei \surnamestart
  Chang\surnameend} \& \bibinfo{author}{Kristofer~SJ \surnamestart
  Pister\surnameend} (\bibinfo{year}{2014}): \emph{\bibinfo{title}{A realistic
  energy consumption model for TSCH networks}}.
\newblock {\sl \bibinfo{journal}{IEEE Sensors Journal}}
  \bibinfo{volume}{14}(\bibinfo{number}{2}), pp. \bibinfo{pages}{482--489},
  \doi{10.1109/JSEN.2013.2285411}.

\bibitemdeclare{article}{zhou2011modeling}
\bibitem{zhou2011modeling}
\bibinfo{author}{Hai-Ying \surnamestart Zhou\surnameend},
  \bibinfo{author}{Dan-Yan \surnamestart Luo\surnameend}, \bibinfo{author}{Yan
  \surnamestart Gao\surnameend} \& \bibinfo{author}{De-Cheng \surnamestart
  Zuo\surnameend} (\bibinfo{year}{2011}): \emph{\bibinfo{title}{Modeling of
  node energy consumption for wireless sensor networks}}.
\newblock {\sl \bibinfo{journal}{Wireless Sensor Network}}
  \bibinfo{volume}{3}(\bibinfo{number}{01}), p.~\bibinfo{pages}{18},
  \doi{10.4236/wsn.2011.31003}.

\end{thebibliography}
\end{document}